\documentclass[%
reprint,
superscriptaddress,
 amsmath,amssymb,
 aps,
]{revtex4-2}

\usepackage{graphicx}
\usepackage{dcolumn}
\usepackage{bm}
\usepackage[colorlinks]{hyperref}
\usepackage{subfigure}

\begin{document}

\preprint{APS/123-QED}

\title{Interplay between multi-spin and chiral spin interactions on triangular lattice}

\author{Li-Wei He}
\affiliation{%
 National Laboratory of Solid State Microstructures and School of Physics, Nanjing University, Nanjing 210093, China
}%

\author{Jian-Xin Li}%
\email{jxli@nju.edu.cn}
\affiliation{%
 National Laboratory of Solid State Microstructures and School of Physics, Nanjing University, Nanjing 210093, China
}%
\affiliation{Collaborative Innovation Center of Advanced Microstructures, Nanjing University, Nanjing 210093, China}

\begin{abstract}
We investigate the spin-$\frac{1}{2}$ nearest-neighber Heisenberg model with the four-site ring-exchange $J_4$ and chiral interaction $J_\chi$  on the triangular lattice by using the variational Monte Carlo method. The $J_4$ term induces the quadratic band touching (QBT) quantum spin liquid (QSL) with only a $d+id$ spinon pairing (without hopping term), the nodal $d$-wave QSL and U(1) QSL with a finite spinon Fermi surface progressively. The effect of the chiral interaction $J_\chi$ can enrich the phase diagram with two interesting chiral QSLs (topological orders) with the same quantized Chern number $C = \frac{1}{2}$ and ground-state degeneracy GSD = 2, namely the U(1) chiral spin liquid (CSL) and Z$_2$ $d+id$-wave QSL. The nodal $d$-wave QSL is fragile and will turn to the Z$_2$ $d+id$ QSL with any finite $J_\chi$ within our numerical calculation. However,  in the process from QBT to the Z$_2$ $d+id$ QSL with the increase of $J_\chi$, an exotic crossover region is found. In this region, the previous QBT state acquires a small hopping term so that it opens a small gap at the otherwise band touching points, and leads to an energy minimum which is energetically more favorable compared to another competitive local minimum from the Z$_2$ $d+id$ QSL. We dub this state as the proximate QBT QSL and it gives way to the Z$_2$ $d+id$ QSL eventually. Therefore, the cooperation of the $J_4$ and $J_\chi$ terms favors mostly the Z$_2$ $d+id$-wave QSL, so that this phase occupies the largest region in the phase diagram.


\end{abstract}

\maketitle


\section{\label{sec1-introduction}introduction}
As an exotic and attractive phase in condensed matter physics, the quantum spin liquid (QSL)\cite{Anderson-RVB-1973, Zhou-rev-qsl-2017, Savary-rev-qsl-2017, Broholm-rev-qsl-2020}  has been studied extensively in recent years. One of the remarkable characteristics of QSL is that it does not possess any magnetic order even at zero temperature. It is in fact not conventional phase triggered by any symmetry breaking at low temperatures obeying the paradigm of Landau's theory, but a new quantum phase with fractional excitations and classified by projective symmetry group (PSG)\cite{Wen-psg-2002}. Intrinsically, this exotic ground state has non-trivial quantum many-body entanglement so that different types of QSLs correspond to different patterns of entanglement. Besides, there is a straight and coarse classification to distinguish two classes of spin liquids based on whether or not there is an energy gap  between the excitation spectrum and ground state. Gapped spin liquids have topological order characterized by the global topological structure and ground-state degeneracy (GSD)\cite{Wen-gsd-1989, Wen-gsd-1990, Wen-gap-qsl-topo-1991}. On the other hand, in gapless systems, the quasiparticle description, such as gapless fermionic (Dirac) spinon,  breaks down\cite{Hermele-Wen-2004}. And they may be characterized by a higher dimensional topological order, or the categorical symmetry\cite{Chatterjee-Wen-arxiv}.

It is expected that a spin system may fall into a QSL instead of a long-range magnetic ordered phase, when quantum fluctuations are strong enough. Usually, spin frustrations, including the geometrical and exchange ones, can enhance effectively the quantum fluctuations. As a celebrated example,  the Kitaev model on the honeycomb lattice\cite{Kitaev-2006} has an exact QSL ground state, where the exchange frustrations arise from the bond-dependent anisotropic spin couplings, though there is no geometrical frustration in this lattice structure.  However, the Kitaev model
is difficulty to realize in a pure spin systems due to its highly anisotropic Kitaev interactions. Recently, many progresses have been achieved to realize the Kitaev interactions  in a class of Mott insulating magnets with strong spin-orbit coupling\cite{Jackeli-Khaliullin-2009, PhysRevB.91.241110, PhysRevB.93.214431, yadav2016kitaev,  PhysRevB.96.115103, PhysRevLett.118.107203, Kitagawa-2018, Kasahara-2018, winter2017models, wen2019experimental, takagi2019concept,sears2020ferromagnetic, li2021identification}.  On the other hand,  there are strong geometrical frustrations in the triangular lattice and kagome lattice, so the searches for the QSL in the materials with these lattice structures are always on the way\cite{Yamashita-organic-Et-tri-qsl-2010, Yamashita-organic-Et-tri-qsl-2010, Han-nat-kagome-qsl-2012, Shimizu-prl-kappaBEDT-2003, Yamashita-np-kappa-BEDT-2008,  Furukawa-nc-kappa-ET-2018, Bordelon-np-tri-qsl-2019,  Yamashita-nc-Et-qsl-2011, Watanabe-nc-Et-qsl-2012, li2015gapless, feng2017gapped,  PhysRevB.98.220409,  He-j4-prl-2018, clark2021quantum, liu2022gapless}.

Traditionally, the QSL is firstly proposed in the triangular antiferromagnetic (AFM) Heisenberg model\cite{Anderson-RVB-1973}.  In recent years, it is generally agreed that the AFM Heisenberg model with only the nearest-neighbour (NN) $J_1$ spin interaction on the triangular lattice exhibits a $120^{\circ}$ magnetic order at low temperatures. Therefore, several possible competing interactions beyond the $J_1$ term are considered. For example, it is found that when the second NN exchange interaction $J_2$ is about $0.08 \alt J_2/J_1 \alt 0.16$, the $120^{\circ}$ order is melted \cite{Iqbal-j1j2-tri-2016} and the QSL would arise \cite{Iqbal-j1j2-tri-2016, Baskaran-j1j2-tri-1989, Jolicoeur-j1j2-tri-1990, Zhu-j1j2-tri-2015, Saadatmand-j1j2-tri-2017, Hu-j1j2-tri-2019}. Nevertheless, there is still doubt about the class of the QSL in this $J_1-J_2$  AFM model,  though Dirac (gapless) spin liquid has been proposed by those very exhausted numerical calculations\cite{Iqbal-j1j2-tri-2016, Hu-j1j2-tri-2019}. This ambiguity is due to  the complication of the possible gapless QSLs, as mentioned above\cite{Chatterjee-Wen-arxiv}. Besides the second NN $J_2$ exchange interaction,  the four-site ring-exchange interaction ($J_4$ term)\cite{Motrunich-j4-2005} was proposed to favor a U(1) QSL with large Fermi surface on the triangular lattice.  It is shown that the $120^\circ$ order is robust against a small $J_4/J_1$ ratio and gives way to the  U(1) QSL at a large one\cite{Motrunich-j4-2005}. In between, several different phases have been claimed, including the chiral $d + id$ QSL, the staggered valence bond solid and nodal $d$-wave QSL,  depending on different effective spin models and numerical techniques\cite{Mishmash-QBT-2013, He-j4-prl-2018, Cookmeyer-j4-prl-2021, Zhao-Liu-j4-prl-2021}.  In particular, a new spin liquid with the  $d + id$ spinon pairing and vanishing hopping has been found when $J_2\approx 0$, whose dispersion exhibits the quadratic band touching at $\vec{k} = 0$, so it is dubbed as QBT spin liquid\cite{Mishmash-QBT-2013}.  This state has been used to explain many of the intriguing experimental properties of the low-temperature phase in organic spin liquid candidate materials, EtMe$_3$Sb[Pd(dmit)$_2$]$_2$\cite{Yamashita-organic-Et-tri-qsl-2010, Yamashita-nc-Et-qsl-2011, Watanabe-nc-Et-qsl-2012} and ${\kappa}$-(BEDT-TTF)$_2$Cu$_2$(CN)$_3$\cite{Shimizu-prl-kappaBEDT-2003, Yamashita-np-kappa-BEDT-2008}.  
We also note that a recent experimental observation of the QSL candidate in inorganic material NaRuO$_2$ with triangular lattice also suggests the importance of the long-range exchange interactions\cite{triangular-np-j4-experiment}, such as the $J_4$ term. On the other hand, the scalar chiral interaction $J_\chi$ that can be derived from the $t/U$ expansion of the Hubbard model at half filling with adding $\Phi$ flux through the elementary triangles \cite{Motrunich-chiral-2006} is introduced. The $J_\chi$ term can naturally stabilize the topological CSLs\cite{Wen-csl-1989} in a spin-$\frac{1}{2}$ Heisenberg model on the triangular lattice\cite{Hu-csl-2016, Wietek-csl-2017}. Consequently, the Berry curvature of CSLs can lead to nontrivial thermal Hall effect\cite{Katsura-thermal-Hall-2010}. Generally, it is expected that this $J_\chi$ interaction could stabilize the $d + id$ chiral spin liquid as induced by the $J_4$ interaction. However, there is a lack of a systematic study of the combined effects of the scalar chiral interaction $J_\chi$  and four-site ring-exchang $J_4$ in the triangular AFM model.

In this work, we investigate the interplay between the scalar chiral interaction $J_\chi$  and four-site ring-exchang $J_4$ in the triangular AFM model with the NN spin interaction $J_1$ using the variational Monte Carlo (VMC) method, focusing on the different chiral spin liquids and their intrinsic topology. We find that a finite $J_\chi$ interaction can stabilize two distinguishable chiral states with time-reversal symmetry breaking,  {\it i.e.},  U(1) CSL with nontrivial fluxes through elementary triangles and chiral Z$_2$ $d + id$-wave QSL. Both of them have the same quantized Chern number $C = \frac{1}{2}$ but actually belong to distinct phases. Especially, there are two different states of the $d + id$-wave phase, which correspond to two energetic minima induced by $J_\chi$. They not only compete with each other, but also happen to be energetic degeneracy within the numerical error under specific conditions. The $120^\circ$ magnetic ordered phase is robust against both a weak four-site ring-exchang and scalar chiral interaction, and is proximate to an algebraic U(1) Dirac QSL by PSG classifications\cite{Zheng-psg-tri-2015, Lu-psg-tri-2016, Bieri-psg-tri-2016, Qi-psg-tri-2018}.  At $J_\chi=0$ and with increase of  $J_4$, the system goes through progressive transitions from the $120^\circ$  magnetic order state to a symmetric Z$_2$ QBT QSL, nodal $d$-wave QSL and U(1) QSL (or called uniform RVB state in literature) with large spinon Fermi surface (U(1) SFS), which is qualitatively consistent with the results reported in Ref.~\cite{Mishmash-QBT-2013}.  We find that both the QBT and nodal $d$-wave QSLs are unstable against a very small chiral interaction $J_\chi$.  The achiral nodal $d$-wave state will immediately give way to the chiral Z$_2$ $d + id$ state with relatively weak spinon pairings. While, the QBT state falls into another chiral Z$_2$ $d + id$ state with a relatively insignificant hopping terms, which is one of the two different $d + id$ states mentioned above. The dispersion of its quasi-particles is almost the same as that of the QBT state, except a very small energy gap in the former. Thereby, we call it the proximate quadratic band touching (PQBT) state. With further increase of $J_\chi$,  the PQBT will enter into the chiral Z$_2$ $d + id$ state.

The paper is organized as follows. In Sec.\ref{sec2-model-method}, we introduce the model, the variational Monte Carlo method and the calculation of Chern numbers by the optimized variational wave functions, mainly focus on the construction of the trial variational wave functions. In Sec.\ref{sec4-result},  we present our results on the phases induced by the $J_4$ and $J_\chi$ terms independently and the effects of their interplay. Section \ref{sec:conclusion} presents a summary. In Appendix A, we introduce the calculation method for the ground-state degeneracy.



\section{\label{sec2-model-method}Model and Method}

\begin{figure}[h]
\centering
\subfigure[]{
    \begin{minipage}[h]{0.9\linewidth}
    \centering
    \includegraphics[width = 1.0\linewidth]{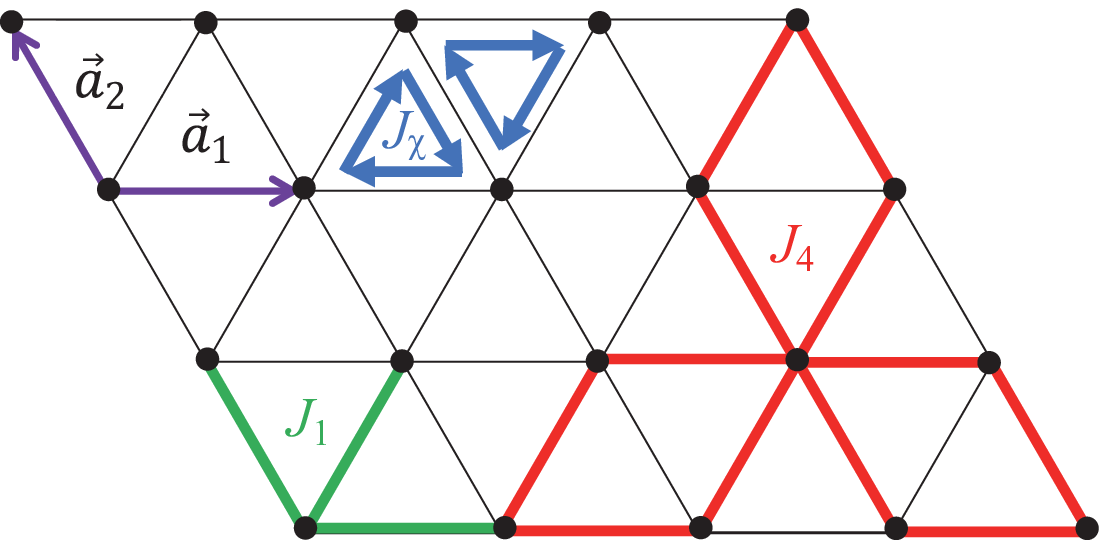}
    \label{fig:model}
    \end{minipage}
    }
\subfigure[]{
    \begin{minipage}[h]{0.44\linewidth}
    \centering
    \includegraphics[width = 1\linewidth]{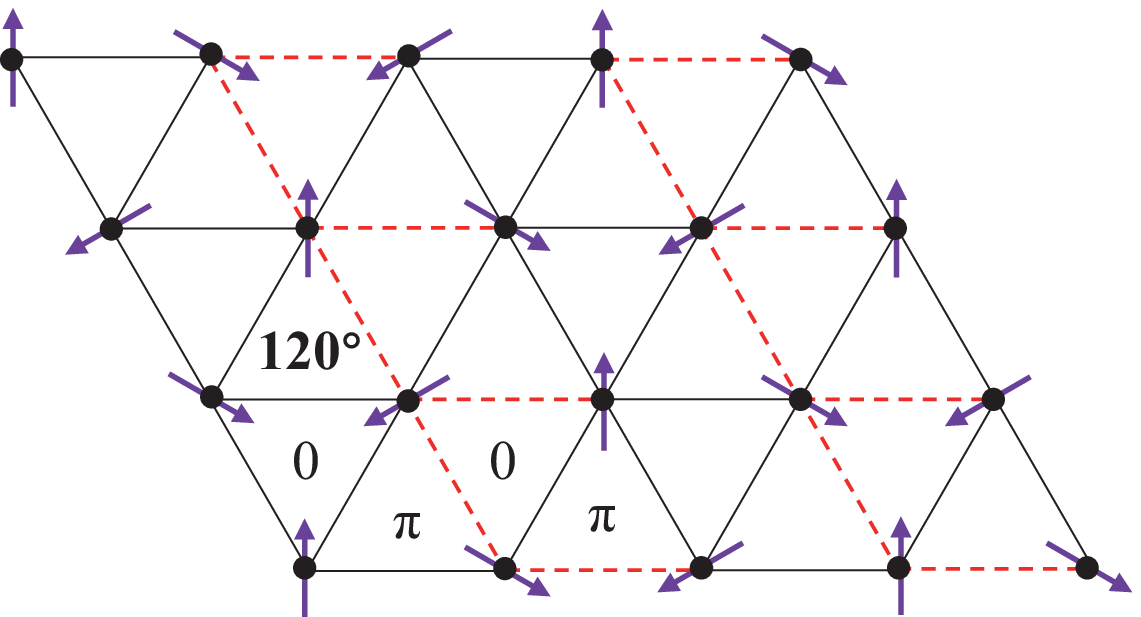}
    \label{fig:120}
    \end{minipage}
    }%
\subfigure[]{
    \begin{minipage}[h]{0.44\linewidth}
    \centering
     \includegraphics[width = 1\linewidth]{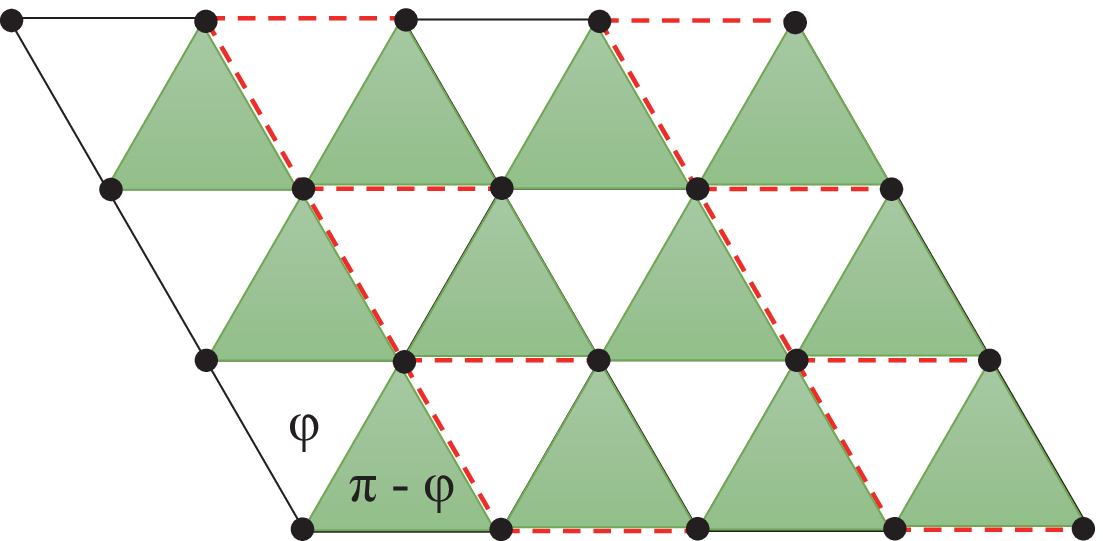}
    \label{fig:csl}
    \end{minipage}
    }
\caption{(a) Triangular lattice with $\vec{a}_1\cdot\vec{a}_2=-\frac{1}{2}$, the lines marked with different colors connect spins involved in the different terms of model (\ref{eq:hamiltonian}). (b) and (c) denotes the ansatzes for the 120$^\circ$ long-range order with alternating 0 and $\pi$ flux, and U(1) CSL with nonzero fluxes through elementary triangles, respectively. The marked red dashed bonds in (b) and (c) denote that the hopping terms along these bonds in $H_\mathrm{mf}$ have the opposite sign compared with those unmarked ones.}
\label{fig:lattice+ansatz}
\end{figure}

The model we considered is written by,
\begin{equation}
    \begin{aligned}
        H =&\  J_1\sum_{\langle i,j \rangle}2\vec{S_i} \cdot \vec{S_j}
            + J_\chi\sum_{i,j,k \in \vartriangle/\triangledown}\vec{S_i} \cdot (\vec{S_j} \times     \vec{S_k})\\&+ J_4\sum_{i,j,k,l \in \lozenge}(P_{ijkl} + \mathrm{H.c.}).
    \end{aligned}
\label{eq:hamiltonian}
\end{equation}
where, the $J_1$ term is the nearest-neighbour (NN) AFM Heisenberg exchange, and the $J_\chi$ term is the scalar chiral interaction with the same magnitude in any elementary triangle (either up triangle $\vartriangle$ or down triangle $\triangledown$, the three sites in the triangle are in clockwise direction, see Fig.~\ref{fig:model}).  The last term $J_4$ is the four-spin coupling  and is given in detail by,
\begin{equation}
    \begin{aligned}
        \sum_{i,j,k,l \in \lozenge}(P_{ijkl} + \mathrm{H.c.}) =&\ 5\sum_{\langle i,j \rangle}\vec{S_i} \cdot \vec{S_j}
            + \sum_{\langle\langle i,j \rangle\rangle}\vec{S_i} \cdot \vec{S_j}\\&+
            4\sum_{i,j,k,l \in \lozenge}[(\vec{S_i} \cdot \vec{S_j})(\vec{S_k} \cdot \vec{S_l}) \\&+ (\vec{S_i} \cdot \vec{S_l})(\vec{S_j} \cdot \vec{S_k}) \\&- (\vec{S_i} \cdot \vec{S_k})(\vec{S_j} \cdot \vec{S_l})] + \frac{1}{4},
    \end{aligned}
\label{eq:four-spin-Hamiltonian}
\end{equation}
where $\langle\langle i,j \rangle\rangle$ denotes the next-nearest neighbor (NNN) bonds and
$i,j,k,l \in \lozenge$ means summing all of elementary four-site rhombi defined by unique NNN pairs
$\langle\langle i,k \rangle\rangle$(see Fig.~\ref{fig:model}). Hereinafter, we set $J_1 = 1$ as the unit of energy.

\begin{figure}[h]
\centering
\includegraphics[width = 1.0\linewidth]{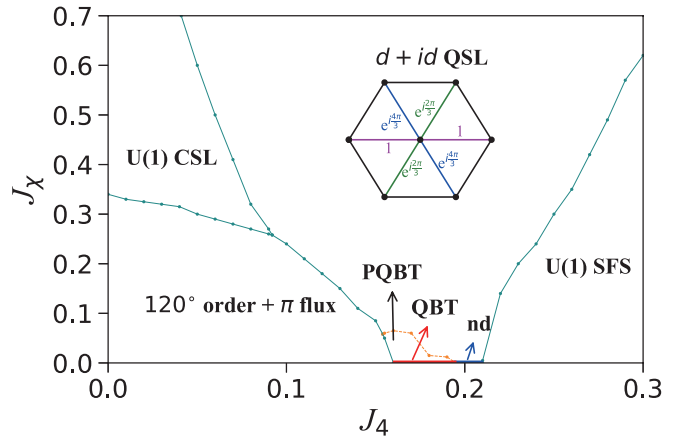}
\caption{Phase diagram of the model Eq.(\ref{eq:hamiltonian}), which includes one $120^\circ$ magnetically ordered phase and the other five disordered phases. The quadratic band touching state (QBT) is marked by red solid line with $J_\chi = 0$. The U(1) QSL with a finite spinon Fermi surface (named as U(1) SFS) localizes in the area with large four-spin interactions. Between them, there is a nodal $d$-wave (nd) disordered state marked by the blue solid line. And the other two phases are chiral spin liquids with TRSB, {\it i.e.}, a U(1) CSL with nonzero flux through the elementary triangles and a chiral $d + id$ QSL with the phases distribution illustrated in the inset. In the area of $d + id$ QSL phase, there is an orange dashed line separating the two states correspodning to the two minima, one of them (we call it PQBT state)  is almost the same as the QBT state except a small gap at the otherwise touching point (see text).}
\label{fig:phase-diagram}
\end{figure}

In this work, we study the phase diagram of the model (\ref{eq:hamiltonian}) with the variational Monte Carlo method. We start from the parton construction (or Abrikosov-fermion spinon representation) of spin $\frac{1}{2}$ operators: $\vec{S} = \frac{1}{2}\sum_{\alpha, \beta = \uparrow,\downarrow}f_{\alpha}^{\dagger}\vec{{\sigma}}_{\alpha\beta}f_{\beta}$. With this representation, we halve one spin to two fermionic spinons.  Hence, the original physical spin-$\frac{1}{2}$ Hilbert space must be recovered by imposing the on-site constraint $\sum_{\alpha}f_{\alpha}^{\dagger}f_{\alpha} = 1$. Furthermore, this fermionic fractionalization will lead to the emergence of a SU(2) gauge structure\cite{Affleck-1988}, which can become explicit if we introduce two doublets,
\begin{equation}
    \psi_1 =
    \left(
    \begin{array}{c}
        f_{\uparrow}\\
        f_{\downarrow}^\dagger
    \end{array}
    \right),
    \psi_2 =
    \left(
    \begin{array}{c}
        f_{\downarrow}\\
        -f_{\uparrow}^\dagger
    \end{array}
    \right),
\end{equation}
and put them into a matrix $\Psi = (\psi_1 \ \psi_2)$, then we rewrite the spin-$\frac{1}{2}$ operator as following:
\begin{equation}
    \vec{S} = \frac{1}{4} Tr\left( \Psi^\dagger \Psi \vec{\sigma}^T \right),
\end{equation}
where $\vec{\sigma}$ is the Pauli matrix.
Besides this formalistic demonstration, intrinsically, the emergent SU(2) gauge structure can also be revealed by the combination of the U(1) gauge structure: $f_\sigma \rightarrow f_\sigma e^{i\alpha}$ and the particle-hole redundancy: $f_\sigma \rightarrow f_\sigma \cos(\beta) + \sigma f_{\bar{\sigma}}^\dagger \sin(\beta)$, both $\alpha$ and $\beta$ are any angles. In addition, it is the unique property for the fermionic representation but not for the bosonic one with only U(1) gauge structure.

Then, we decouple the Hamiltonian Eq. \ref{eq:hamiltonian} into a general quadratic fermionic Hamiltonian of the form with a unconsidered constant number,
\begin{equation}
    \begin{aligned}
    H_{\mathrm{mf}} =& \sum_{i,j}\left( t_{ij}f_{i\sigma}^\dagger f_{j\sigma} + \Delta_{ij}f_{i\uparrow}^\dagger f_{j\downarrow}^\dagger + \mathrm{H.c.} \right) \\
    &+ 2\sum_{i}\vec{M}_i \cdot \vec{S}_i + \mathrm{const},
    \end{aligned}
\label{eq:mf-hamiltonian}
\end{equation}
where an additional background field $\vec{M}_i$ is introduced to induce a static magnetic long-range order as done before~\cite{Iqbal-j1j2-tri-2016, liu-prl-2018-honeycomb, Zhao-Liu-j4-prl-2021}.
Because the model we considered is very complicate,  we will get so many mean-field variational parameters if we consider all channels. As well known, the numerical simulations with so plenty of variational parameters are almost not reliable and usually make physics unclear in a limited time cost. Combining the previous works Ref.~\cite{Grover-j4-prb-2010, Mishmash-QBT-2013, He-j4-prl-2018, Cookmeyer-j4-prl-2021, Zhao-Liu-j4-prl-2021, Hu-csl-2016, Bieri-psg-tri-2016} with the physics we focus on, here we only consider various NN-bond ($\langle i,j \rangle$) hoppings $t_{\langle ij \rangle}$ and pairings $\Delta_{\langle ij \rangle}$ with the background field $\vec{M}_i$.

After constructing the ground state $|G\rangle_{\mathrm{mf}}$ of the mean-field Hamiltonian Eq.\ref{eq:mf-hamiltonian}, we utilize the Gutzwiller projective operator $P_G = \prod_i(1 - n_{i\uparrow}n_{i\downarrow})$ to $|G\rangle_{\mathrm{mf}}$ to enforce the local particle number constrain: $n_{i\uparrow} + n_{i\downarrow} = 1$. Finally, we obtain a general trial variational wave function $|\Psi(\mathcal{P})\rangle = P_G|G\rangle_{\mathrm{mf}}$, where $\mathcal{P}$ denotes variational parameters. For those states without magnetic order, we can set $\sum_{i}S_i^z=0$ (or $N_\uparrow = N_\downarrow = N/2$, N is number of the lattice) without loss of generality. We fix the spinon chemical potential $\mu = t_{ii}$ such that $|G\rangle_{\mathrm{mf}}$ is at half filling before projection\cite{Mishmash-QBT-2013}.  Actually, for those ansatzes without spinon pairings, we just need the N occupied states of $H_{\mathrm{mf}}$ to construct the trial wave functions, so the chemical potential makes no difference for this procedure so that we can abandon it as a variational parameter in practical numerical calculations. As a result, the rest parameters $\mathcal{P}=(t_{\langle ij \rangle}, \Delta_{\langle ij \rangle}, \vec{M}_i)$ in the mean-field Hamiltonian are used as variational parameters. And we consider different types of ansatzes, including various Z$_2$, U(1) QSLs, and $120^\circ$ AFM ordered states to construct the initial trial wave functions, where the parameters $\mathcal{P}$ are optimized by minimizing the trial energy $E(\mathcal{P}) = \langle\Psi|H|\Psi\rangle/\langle\Psi|\Psi\rangle$. We adopt a triangle lattice with torus geometry: $L_1 = L_2 = 12$ ($L_{1,2}$ are the lengths along the two reciprocal basic vectors ($\vec{a}_{1,2}$) of primitive cell, see Fig.~\ref{fig:model}).

To extract the topological properties for these gapped CSLs, we calculate Chern numbers by use of the optimized variational wave functions $|\Psi\rangle_{\mathrm{opt}}$ with twist boundary condition\cite{Hu-csl-2016, Sheng-chern-num-2003}, as following
\begin{subequations}
\label{eq:chern-bp}
\begin{equation}
    f_{i+L_k,\uparrow} = f_{i,\uparrow}e^{i\theta_k}; f_{i+L_k,\downarrow} = f_{i,\downarrow}e^{-i\theta_k}(k = 1,2),
    \label{eq:twist-bc}
\end{equation}
\begin{equation}
    {{\rm BP}(p)} = {\rm Im} \left( \ln \prod_{i=1}^{4}\langle\Psi^{{p}_{i+1}}|\Psi^{{p}_{i}}\rangle \right),
    \label{eq:berry-phase}
\end{equation}
\begin{equation}
   {C}_{\mathrm{total}}= \frac{1}{2\pi}\sum_{p} {{\rm BP}(p)},
    \label{eq:chern-num}
\end{equation}
\end{subequations}
where Eq.~\ref{eq:twist-bc} expresses the twist boundary condition, BP($p$) is the Berry phase in the plaqutte $p$, the label $i=1,2,3,4$ denotes the four corners of the $p$th plaqutte, and the overlaps are calculated by Monte Carlo method. Eq.~\ref{eq:chern-num} is used to calculate the Chern numbers numerically. In our calculations, we have checked the results with the numbers of mesh plaquttes, $N_p = 36, 64, 100, 144$, and find that $N_p = 100$ is large enough so that the Chern numbers do not change by further increasing the mesh plaqutte size.

\begin{figure*}
\subfigure[]{
    \begin{minipage}[h]{0.22\linewidth}
    \centering
     \includegraphics[width = 1\linewidth]{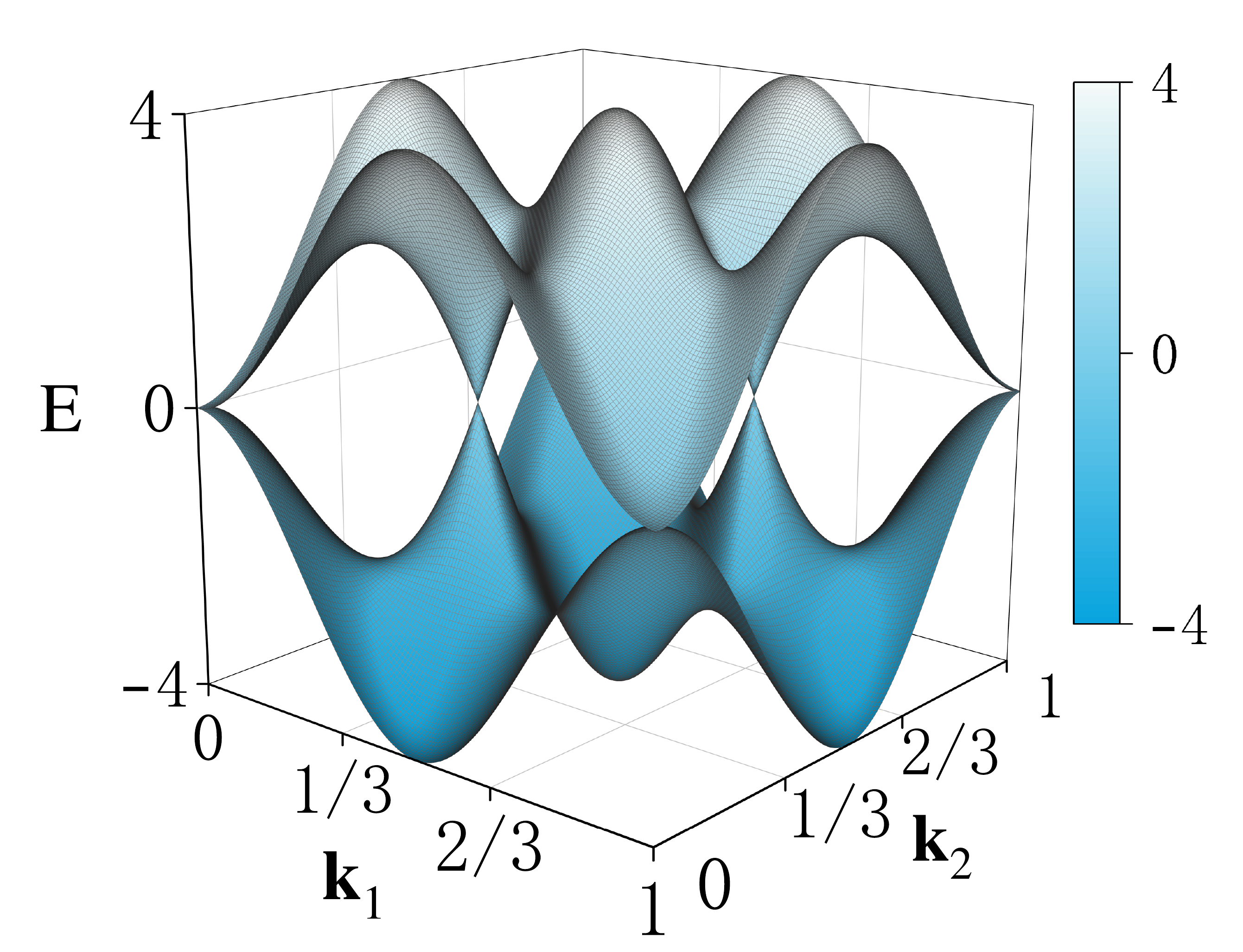}
    \label{c}
    \end{minipage}
    }%
\subfigure[]{
    \begin{minipage}[h]{0.22\linewidth}
    \centering
    \includegraphics[width = 1\linewidth]{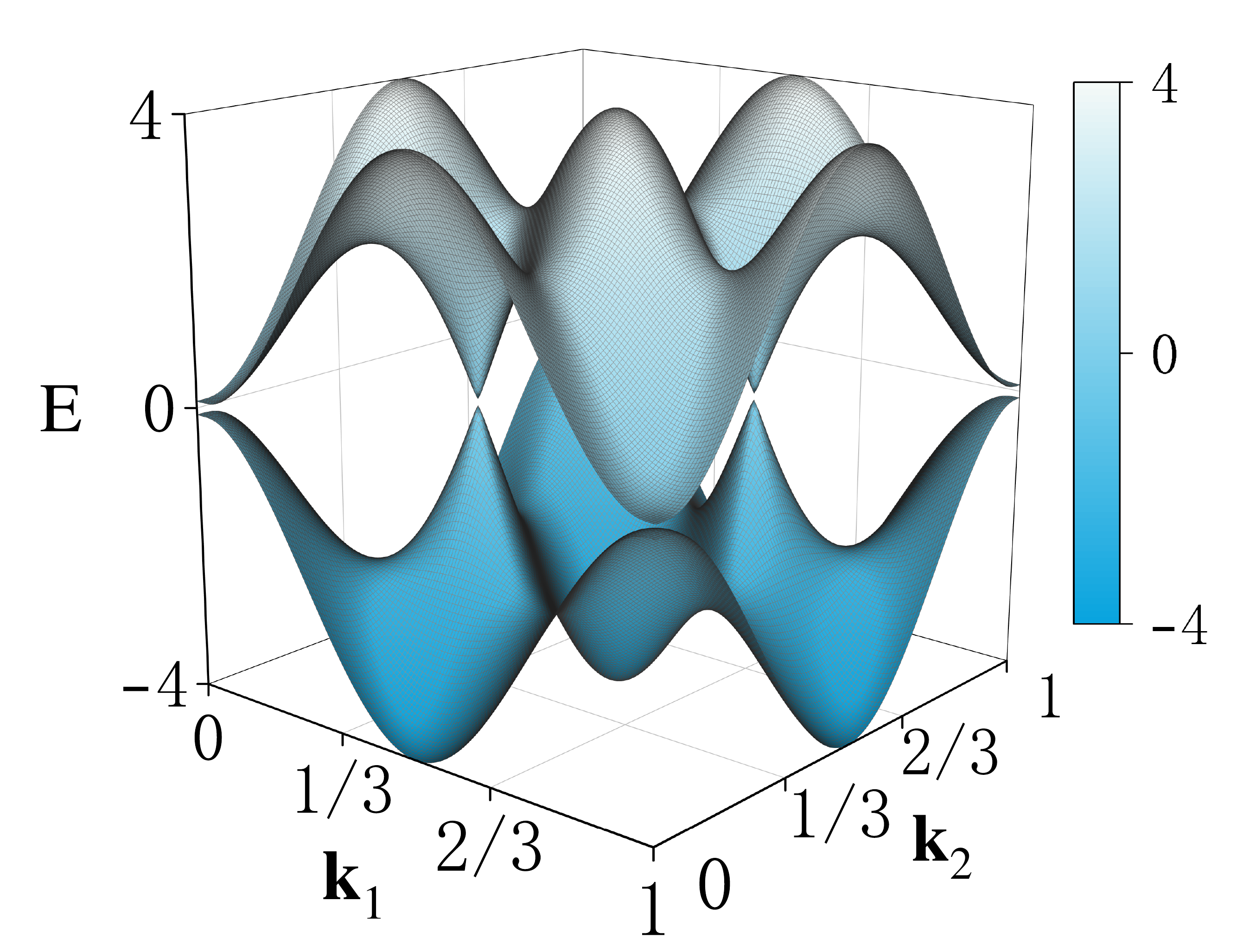}
    \label{fig:band-PQBT}
    \end{minipage}
    }%
\subfigure[]{
    \begin{minipage}[h]{0.22\linewidth}
    \centering
    \includegraphics[width = 1\linewidth]{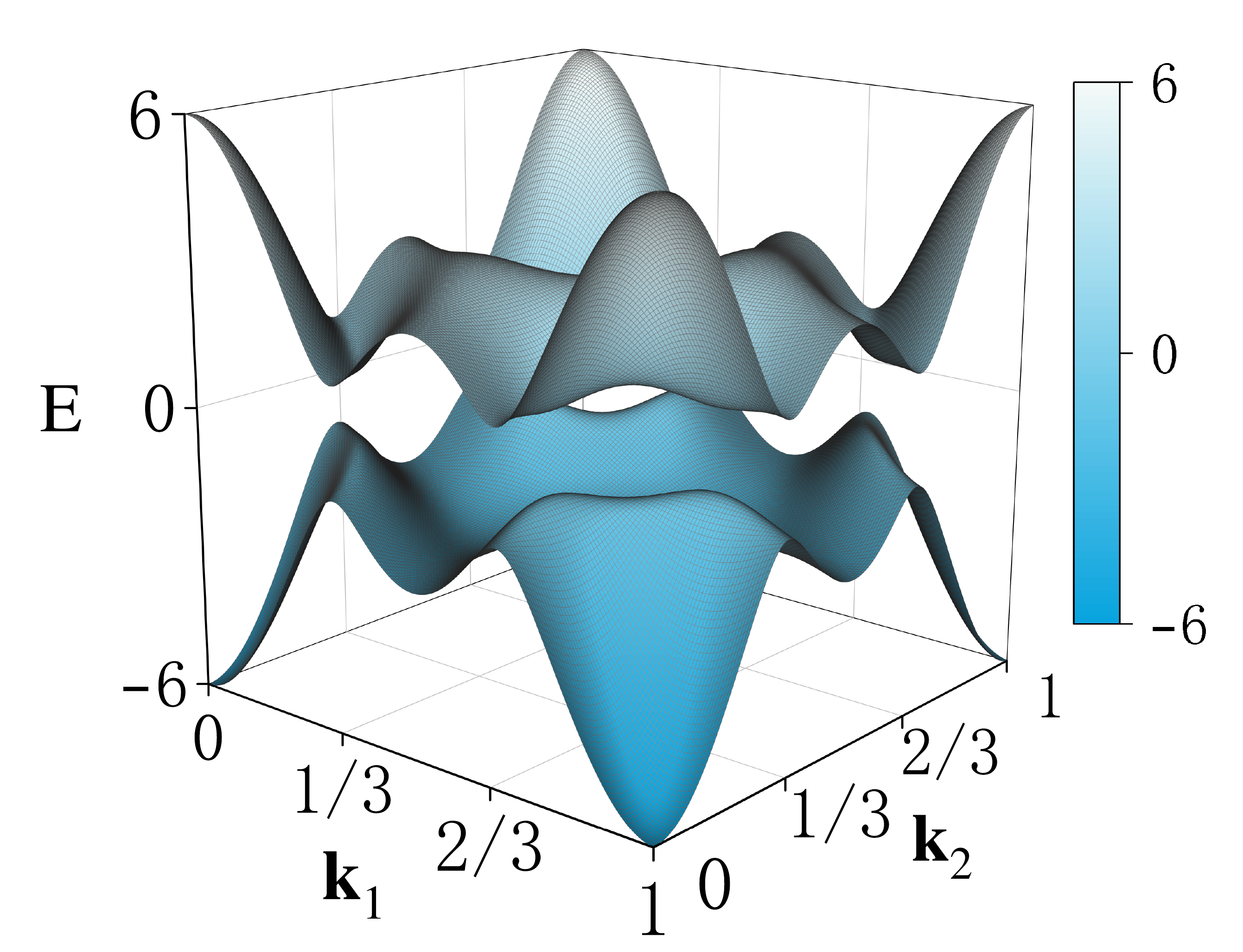}
    \label{fig:band-did}
    \end{minipage}
    }%
\subfigure[]{
    \begin{minipage}[h]{0.27\linewidth}
    \centering
    \includegraphics[width = 1\linewidth]{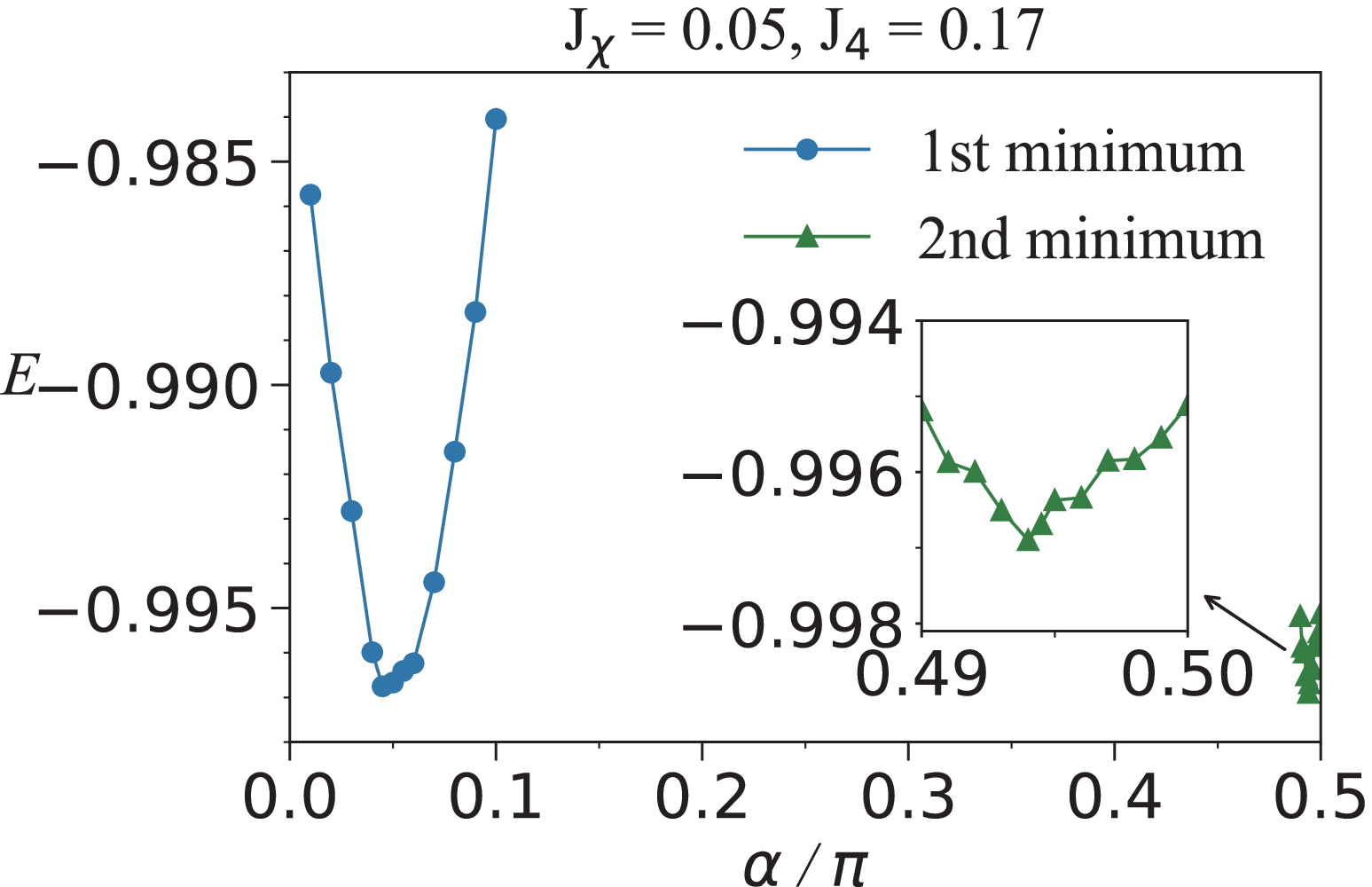}
    \label{fig:minima-vs-alpha}
    \end{minipage}
    }
\caption{(a), (b) and (c) are dispersions of the quasiparticles of the QBT state, PQBT and $d + id$ state, respectively. These dispersions are used only for exhibiting the differences among the three QSL states at the mean-field level, and the energy `E' is chosen in arbitrary unit. (d) exhibits two minima in the PQBT state with $J_\chi = 0.05$ and $J_4 = 0.17$ (see text). The inset is used for an enlarged view of the variational energy of the 2nd minimum.}
\label{fig:band}
\end{figure*}

\begin{figure}[h]
\centering
\includegraphics[width = 1.0\linewidth]{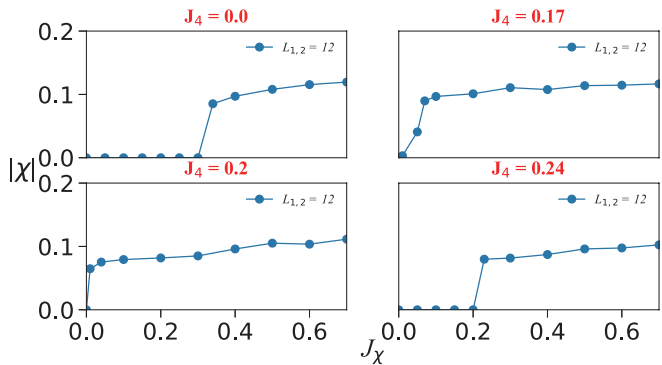}
\caption{Chiral order parameter $|\chi| = |<\vec{S_i} \cdot \vec{S_j} \times \vec{S_k}>|$ ($i$, $j$ and $k$ are the three vertices of each elementary triangle) as a function of the chiral interaction $J_\chi$, for different four-site ring-exchange interactions $J_4 = 0.0, 0.17, 0.2, 0.24$. It is averaged over all triangles of the lattice.}
\label{fig:chirality}
\end{figure}

\section{\label{sec4-result}Results}

Let us firstly list all six phases we find, which are summarized in the phase diagram shown in Fig.~\ref{fig:phase-diagram}. It consists of one long-range magnetic $120^\circ$ order and five disordered phases, the later includes two chiral spin liquids with nontrivial Chern number and three achiral ones. We have considered the potential tetrahedral order\cite{Cookmeyer-j4-prl-2021, Wietek-csl-2017, Zhang-tetra-2021}, especially, for a large chiral interaction $J_\chi/J_1$ at which the spins may behave as classical objects. But, we find that this ordered phase is not energetically favored in the parameter range we considered using our variational numerical simulations, compared with the phases we find in the phase diagram.  In the following, we will discuss these found phases in detail.


\subsection{\label{sec4-2-j_4} Effects of $J_4$}

In this subsection, we fix $J_\chi = 0$ to investigate the effects of the four-site ring-exchange interaction $J_4$.  As well known, the $120^\circ$ long-range magnetically ordered phase exists when $J_4=0$. In fact, as has been shown before~\cite{Iqbal-j1j2-tri-2016,Zhao-Liu-j4-prl-2021} , a background magnetic field $\vec{M}_i = |\vec{M}|(\cos(\vec{Q}\cdot\vec{r}_i),\sin(\vec{Q}\cdot\vec{r}_i),0)$ is needed to induce this order in the framework of the variational Monte Carlo method, where $|\vec{M}|$ denotes the field amplitude and $\vec{Q} = (1/3, 1/3)$ (here we adopt the single-$\vec{Q}$ approximation and the reciprocal bases of the primitive cells as denoted in Fig.~\ref{fig:120}). Otherwise, this magnetic phase will disappear and fall into a U(1) Dirac QSL yielded by the NN-bond hopping terms containing alternate 0 and $\pi$ flux in the elementary triangles.  So, we label it as `120$^\circ$ {\rm order}+$\pi$ flux' in the phase diagram Fig.~\ref{fig:phase-diagram}.

With the introduction of the $J_4$ term, we find that the $120^\circ$ order can survive in an extended region up to  $J_4\approx 0.153$. Then, it enters into the QBT state with only a $d + id$ spinon pairing, but without the hopping term as has been found before\cite{Mishmash-QBT-2013, Bieri-psg-tri-2016}. The phases on different bonds of the $d + id$ spinon pairing state are illustrated in the inset of Fig.~\ref{fig:phase-diagram}. This QBT state exhibits the quadratic band touching at $\vec{k} = 0$  in its dispersion, as shown in Fig.~\ref{fig:band}(a). With the further increase of $J_4$ term, the Z$_2$ nodal $d$-wave state will be competitive and become the ground state [Fig.~\ref{fig:phase-diagram}]. This state is characterized by a singlet spinon pairing with the pairing function $\Delta_{\langle ij \rangle} = \Delta_{\langle ji \rangle}$ and its magnitude $\Delta_{{\rm nd}} = {\rm Re}(\Delta_{d+id})$, and a bond independent hopping term. The identification of this state is qualitatively consistent with that in Ref.\cite{Mishmash-QBT-2013}, where it also exists in a significant area in the phase diagram. While, it has been argued that the nodal $d$-wave QSL is not energetically favored with the increase of system size and is not the ground state in the thermodynamic limit\cite{Zhao-Liu-j4-prl-2021}. We suggest two possible reasons for this difference. One is the detail form of the $J_4$ term, and we adopt the same form as Ref.~\cite{Mishmash-QBT-2013}. Another one is that the stability of this gapless state is sensitive to the lattice geometry. If we use the torus geometry, such as $L_{1,2} = 12$, we always suffer dilemmas about the construction of trial many-body wave function because there are plenty of normal states for quasi-particles in the nodal $d$-wave state in the thermodynamic limit. In fact, it is better to use the lattice size with $L_1 \neq L_2$ (such as $L_1=10$, $L_2=11$) instead of the torus geometry in practical VMC procedures, as has been noticed before\cite{Grover-j4-prb-2010}. The pairing amplitude $|\Delta_{\rm nd}|$ will fade away as the four-spin term $J_4$ increases. Finally, a U(1) SFS state will emerge and extends to the largest $J_4$ we considered, as shown in Fig.~\ref{fig:phase-diagram}.

\subsection{\label{sec4-1-j_chi}Effects of $J_\chi$}
As the chiral interaction term $J_\chi$ breaks the time-reversal symmetry, it is expected that it can induce or stabilize some chiral phases. Starting from the $120^\circ$ long-range magnetically ordered phase, we find that it is robust against a chiral interaction $J_\chi<0.34$. In fact, it is found that the strength $|\vec{M}|$ of the $120^\circ$ order decreases a little with the increase of $J_\chi$, as we have checked with the lattice size $L_{1,2}=6$, 12, 18 and 24. In this region, the chiral order parameter $|\chi| = |<\vec{S_i} \cdot \vec{S_j} \times \vec{S_k}>|$ defined as the expectation averaged over all triangles ($i$, $j$ and $k$ are the three vertices of each elementary triangle) is also found to be zero. When $J_\chi \agt 0.35$, the CSL phase will be energetic favorite and become dominant, as shown in Fig.~\ref{fig:phase-diagram} where the boundary between $120^\circ$ order and CSL is determined by comparing the energy of the two phases. Associated with this transition, the chiral order parameter $|\chi|$ shows a step-like rise from zero to a finite value as shown in Fig.~\ref{fig:chirality} ($J_4=0$). Compared to the result obtained with only the $J_4$ term in the last subsection, we suggest that the $120^\circ$ magnetic order is more stable against the chiral interaction.

This CSL state is not only time-reversal symmetry breaking but also lattice reflection symmetry breaking. However, the combination of the two symmetries is preserved\cite{Bieri-psg-tri-2016}. In this state, there exist alternating $\psi$ and $\pi-\psi$ fluxs through the down triangles and up triangles without spinon pairings (see Fig.~\ref{fig:csl}). We find that the flux $\psi$ increases as $J_\chi$ increases in the whole area of the CSL state in the phase diagram, consequently it leads to an increase of the energy gap of quasiparticles. According to our calculation, the gap reaches its maximum when $\psi = \frac{\pi}{2}$. It corresponds to the case $\psi=\pi-\psi=\frac{\pi}{2}$ so that the fluxes distribute uniformly along all triangles. Hence, the chiral interaction tends to make the alternating fluxes be homogeneous and increases the gap. As denoted in Fig.~\ref{fig:120}, the 120$^\circ$ order+$\pi$ flux phase has alternating $0$ and $\pi$ fluxs through the down triangles and up triangles. So, the CSL state inherits this flux structure and acquires a finite $\psi$ to yield gap compared with gapless Dirac spin liquid($\psi$ = 0). We note that, when we get an optimized flux $\psi_{\mathrm{opt}}$ by VMC, simultaneously, another CSL with $\psi^{'}=\pi-\psi_{\mathrm{opt}}$ is degenerate with it in the thermodynamic limit, suggesting the existence of a gauge degree of freedom. Though these nonzero fluxes $\psi$s are not equal within different areas in the CSL phase, the corresponding states exactly belong to the same type of QSLs according to the PSG classification. In other words,  any two different states of this CSL phase with different fluxes can be interconverted by a local unitary operation without gap closing\cite{Chen-Gu-Wen-2010}. More intrinsically, they possess the same topological structure protected by the sharing PSG. So, the CSLs with different $\psi$s have the same total Chern number $C_{\mathrm{total}}=2$ with high accuracy as shown in Fig.~\ref{fig:bp-csl}. Finally, we must emphasize the total Chern number in Eq.~\ref{eq:chern-num} includes two kinds (spin-up and spin-down) of spinons and two periods for the spin operators, which results in a fractional quantized Chern number $C = \frac{1}{2}$\cite{Hu-csl-2016, hu-prb-kagome-2015}.

\subsection{\label{sec4-3-interply}Interplay between $J_\chi$ and $J_4$}

\begin{figure*}
\subfigure[]{
    \begin{minipage}[h]{0.14\linewidth}
    \centering
    \includegraphics[width = 1.0\linewidth]{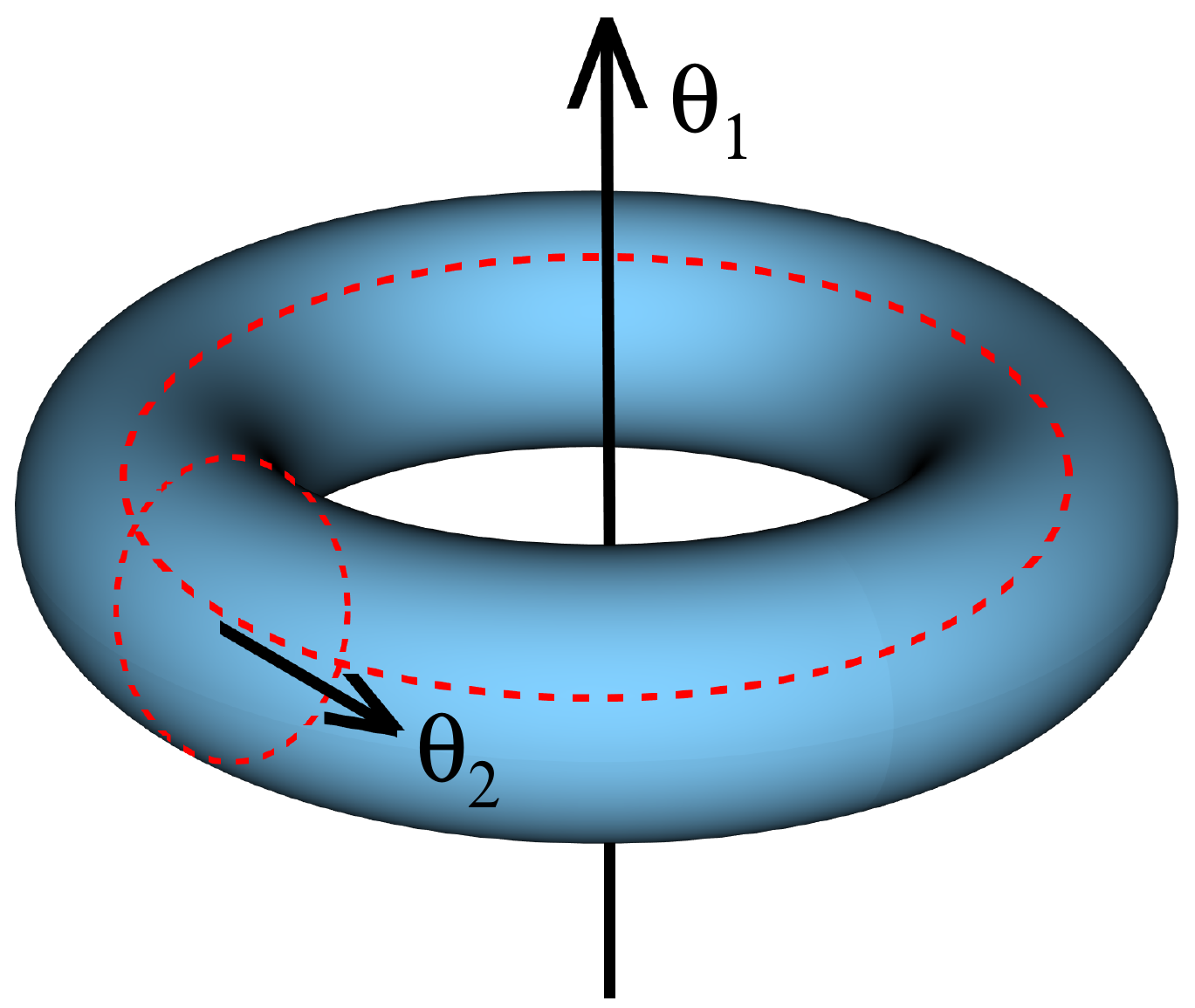}
    \label{fig:torus}
    \end{minipage}
    }%
\subfigure[]{
    \begin{minipage}[h]{0.27\linewidth}
    \centering
    \includegraphics[width = 1\linewidth]{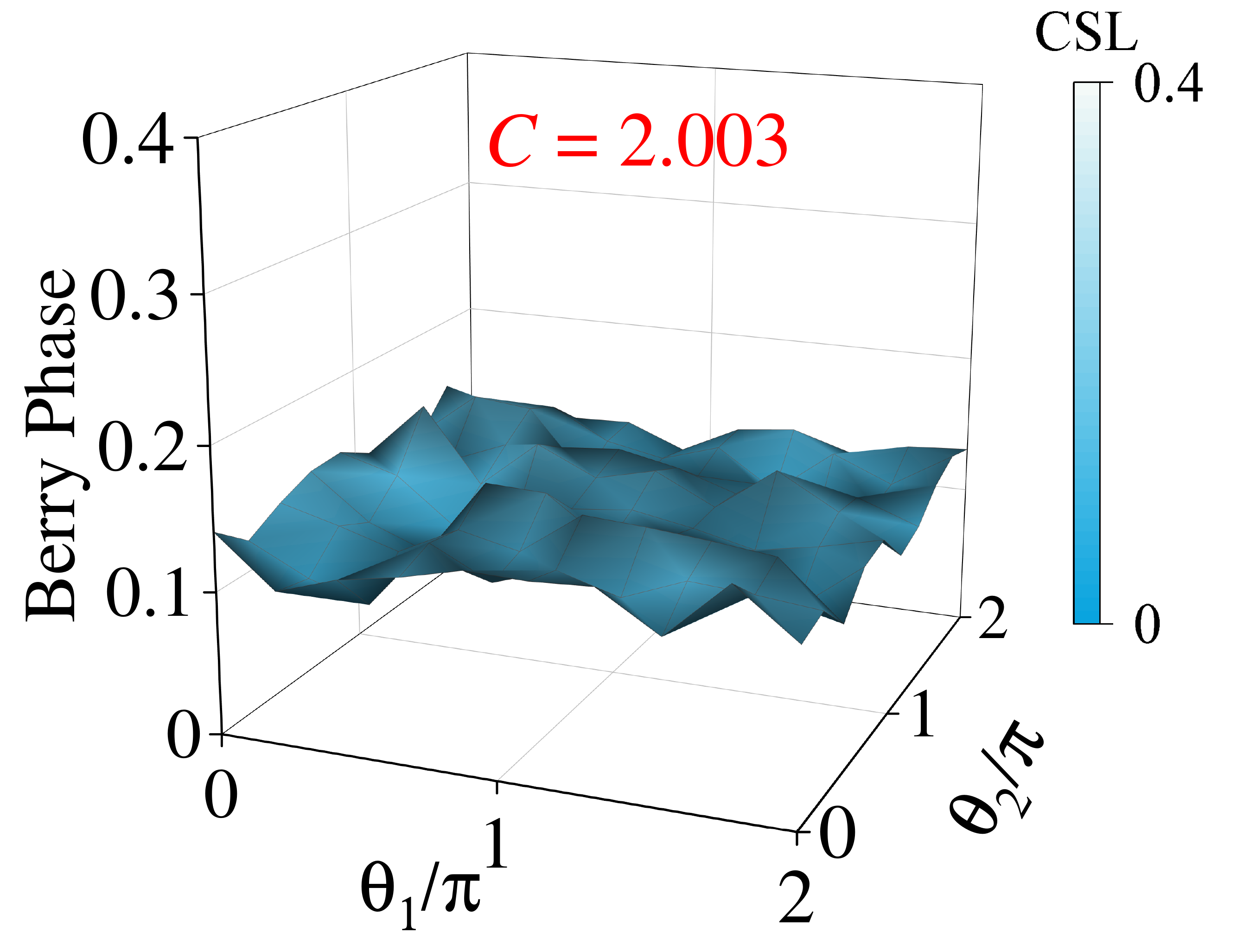}
    \label{fig:bp-csl}
    \end{minipage}
    }%
\subfigure[]{
    \begin{minipage}[h]{0.27\linewidth}
    \centering
    \includegraphics[width = 1\linewidth]{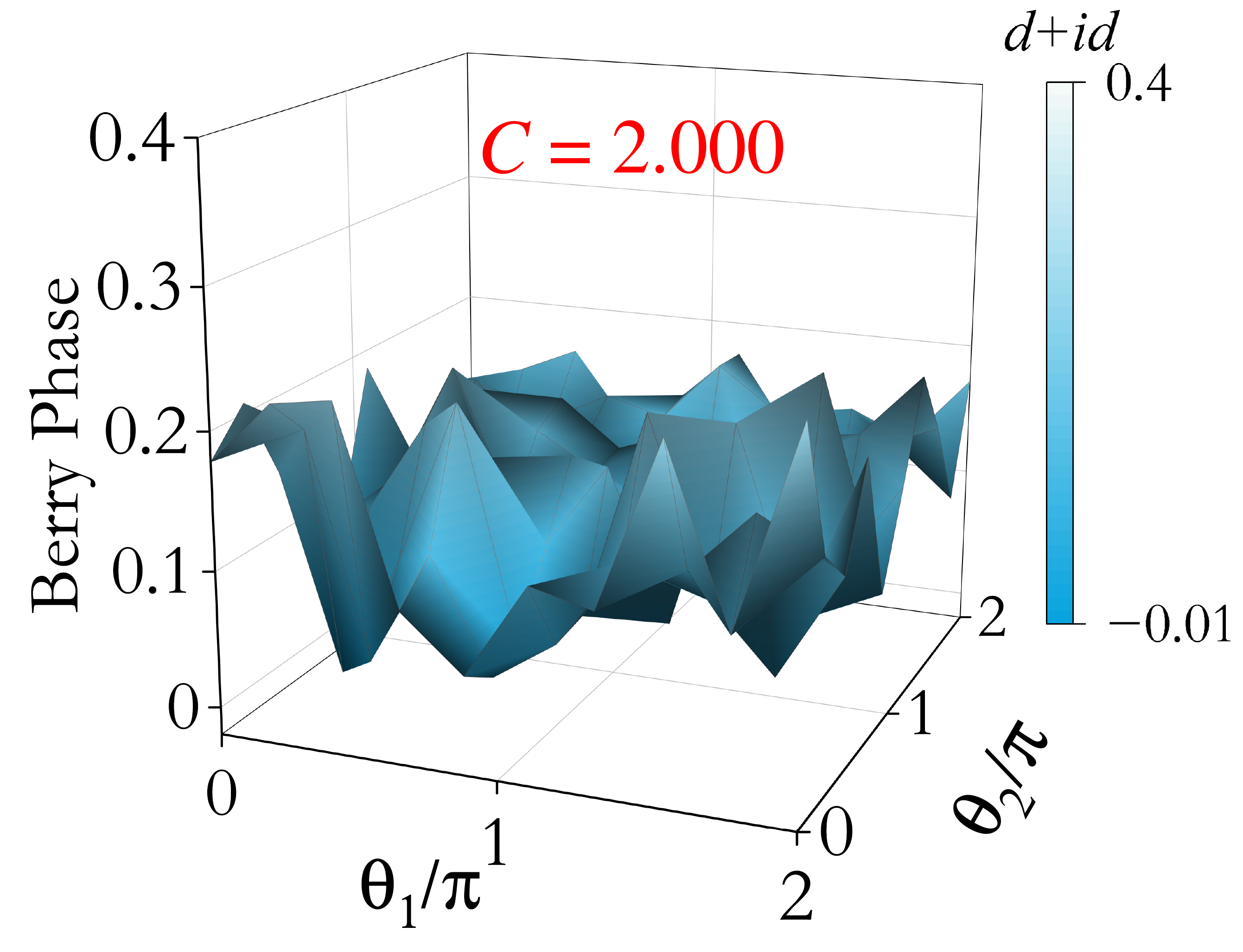}
    \label{fig:bp-did}
    \end{minipage}
    }%
\subfigure[]{
    \begin{minipage}[h]{0.27\linewidth}
    \centering
    \includegraphics[width = 1\linewidth]{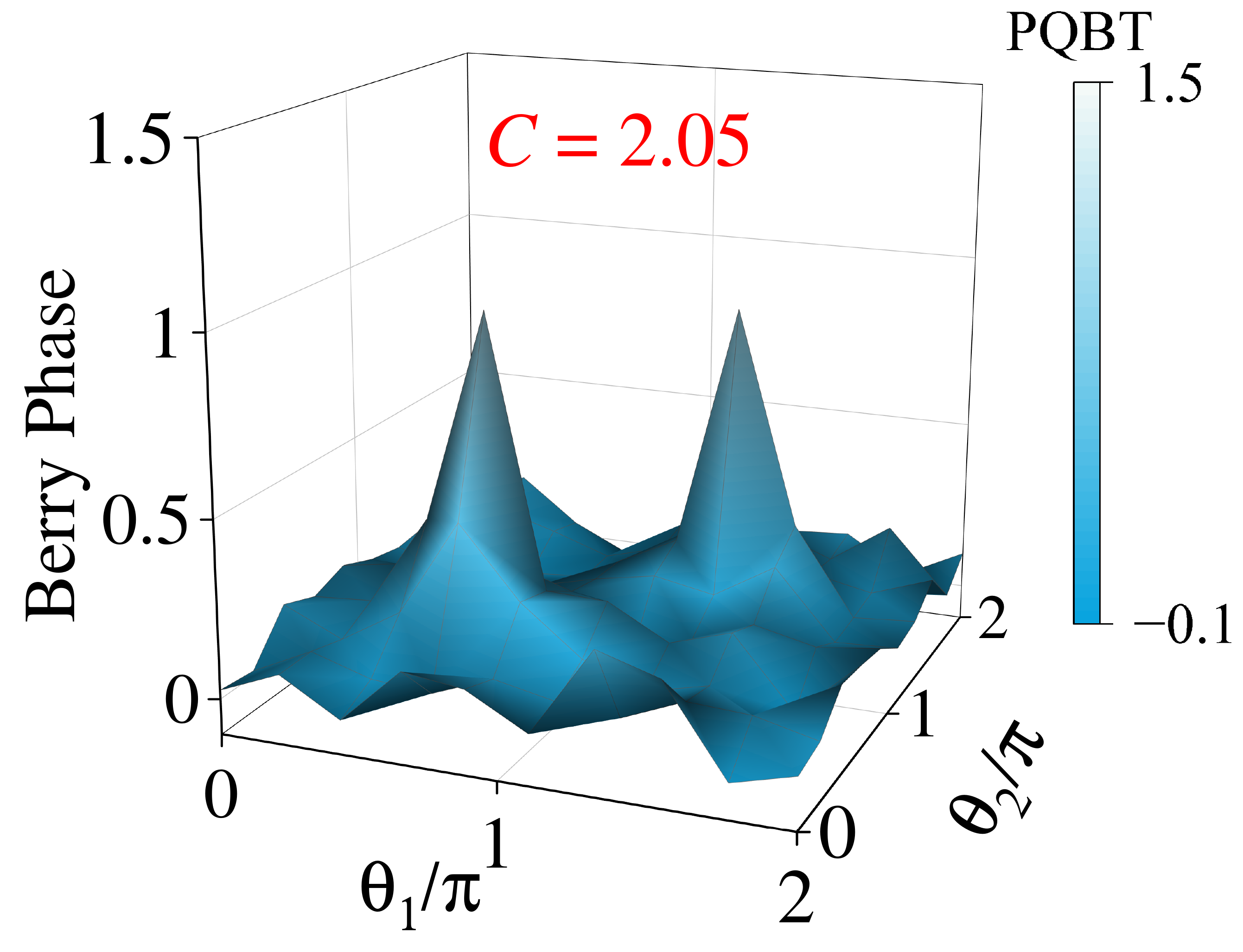}
    \label{fig:bp-PQBT}
    \end{minipage}
    }
\caption{(a) A torus is generated by the tow red dashed rings, where $\theta_1$ and $\theta_2$ are fluxes through the two holes. We compact $L_1 = L_2$ triangular lattice to this torus and mesh the phase space to 100 plaquttes to calculate the Chern number. (b), (c) and (d) present the distribution of the Berry phases in the $\theta_1$-$\theta_2$ plane for three chiral QSLs, $U(1)$ CSL, $d + id$ and PQBT, respectively. The total Chern number is marked with red numbers in each figure.}
\label{fig:topologe}
\end{figure*}

Let us first look at the effects of the four-site ring-exchange interaction on the CSL induced by the chiral spin interaction. The U(1) CSL without spinon pairings is stability against a small  $J_4$ interaction, but will transit into the chiral Z$_2$ $d+id$-wave QSL state with the increase of $J_4$.  This chiral state is not only TRSB but also lattice reflection symmetry breaking, but the combination of the two symmetries is preserved\cite{Bieri-psg-tri-2016} as the same as U(1) CSL. The $d+id$ QSL has the characteristic spinon pairing function whose phases show different distributions on different bonds as illustrated in the inset of Fig.~\ref{fig:phase-diagram}, but its hopping term is uniform along different bonds.  As discussed above, the $\psi$ in the alternating $\psi$ and $\pi-\psi$ fluxes in the CSL state increases with $J_\chi$, and tends to a homogeneous distribution of fluxes. This relative homogeneous distribution is more compatible with the uniform distribution of the hopping term in the $d+id$ QSL, which might be beneficial to the transition into the $d+id$ QSL. Hence, the region of the CSL state in the phase diagram becomes narrow with $J_\chi$, as can be clearly seen in Fig.~\ref{fig:phase-diagram}.

Then, we turn to discuss the instability with respect to  the $J_\chi$ term of the four phases found with $J_4\neq 0$ in the Sec.~\ref{sec4-2-j_4}.  We find that the two QSLs with gap nodes in the spinon pairing function are fragile to the chiral spin interaction, while the U(1) SFS state is more stable. Noticeably, all the three QSLs will eventually give way to the same phase, namely the $d+id$ QSL. For the $120^\circ$ magnetically ordered phase, it also transits into the $d+id$ QSL when $J_4$ exceeds the triple point value in the phase diagram. Due to the existence of a large spinon Fermi surface in the U(1) SFS, a notable $J_\chi$ is needed to destabilize it, for example,  $J_\chi \approx 0.6$ is required for $J_4=0.3$. Hence, the U(1) SFS is favored and stabilized by large $J_4$ terms. In the meantime, the region of the U(1) SFS is also narrowed by $J_\chi$. Therefore, we find that the $d + id$  QSL occupies the largest region in the phase diagram with the cooperation of the $J_\chi$ and $J_4$ interactions [see Fig.~\ref{fig:phase-diagram}]. As mentioned in Ref.~\cite{Mishmash-QBT-2013}, the third-neighbor AFM Heisenberg interaction is also suggested to be able to expand the insignificant region of the chiral $d + id$ QSL state. Our results show that the chiral interaction is an alternative to accomplish that.

Now, let us further discuss the detail of the instabilities of the two QSLs with gap nodes (nodal $d$-wave and QBT) with respect to the chiral interaction. The nodal $d$-wave QSL is found to be unstable into the $d+id$ QSL with any finite $J_\chi$ with our numerical calculations, so its region in the phase diagram in Fig.~\ref{fig:phase-diagram} is denoted as a blue line. When the chiral interaction $J_\chi$ is turned on, the QBT state with gapless quadratic band touching also opens gaps at the quadratic touching point $\vec{k} = 0$ and another two Dirac points ($ \boldsymbol{K} = \left(1/3, 1/3\right)$, $\boldsymbol{K}^{'} = \left(2/3, 2/3\right)$, here we adopt the reciprocal bases of the primitive cells as denoted in Fig.~\ref{fig:120}) by acquiring a nonzero hopping term. However, Fig.~\ref{fig:band} shows that its quasiparticle dispersion is quite different from that of the chiral $d+id$ QSL as mentioned above. On the other hand, though its dispersion looks to be quite similar to that of the QBT [see Fig.~\ref{fig:band}], it has small gaps at those $\vec{k}$ points, so its low temperature properties will show differences with the QBT. Therefore, we dub it the proximate QBT, though it shares the same PSG with the chiral Z$_2$ $d+id$ QSL. In fact, we find a strong competition between the PQBT state and the Z$_2$ $d+id$ state in this region, and it exhibits as the existence of two local minima in the energy curve corresponding to these two states. To characterize the two states, we define a parameter of $\alpha =\arctan (\left| \Delta \right|/t)$, where $\left| \Delta \right|$ represent the amplitude of spinon pairing term and $t$ is the hopping one. A typical energy curve as a function of $\alpha$ for $J_\chi=0.05$ and  $J_4=0.17$ is presented in Fig.~\ref{fig:minima-vs-alpha}. The minimum at $\alpha=0.495\pi$ comes from the PQBT state, which deviates slightly from the value $0.5\pi$ of the QBT state after acquiring a small nonzero $t$. And another one at $\alpha=0.05\pi$ corresponds to $d+id$ QSL state. For $J_\chi=0.05$ and $J_4=0.17$, the local minimum at $\alpha=0.495\pi$ has a lower energy, so the PQBT state is energetically favorable. Thus, we show that the chiral interaction induces two stable states when starting from the QBT, one is the PQBT and the other is the $d+id$ QSL. Firstly, the PQBT state has a lower energy, but the energy difference between them is reduced with $J_\chi$. So, the two state will have the same energy at the critical value. We collect these critical values and plot them as an orange dashed line in the phase diagram Fig.~\ref{fig:phase-diagram}.  Above this line, the system enters into the $d+id$ QSL.  We note that there is another difference between these two states, {\it i.e.},  the chiral order parameter $|\chi|$ approaches to the saturate value after crossing this line, but increases smoothly in the crossover PQBT region, as can be seen in Fig.~\ref{fig:chirality}. While, it has a step-like rise when starting from the nodal $d$-wave state as shown with $J_4=0.2$, or from the U(1) SFS with $J_4=0.24$.

When turning on the spin chiral interaction,  one will expect it induces nonzero Chern numbers. To calculate the Chern number, we compact the triangular lattice on the torus with  $\theta_1$ and $\theta_2$ the fluxes through the two holes [see Fig.~\ref{fig:torus}]. We find nonzero Berry phases as a function of $\theta_1$ and $\theta_2$, which are presented in Fig.~\ref{fig:topologe}. For the three chiral states including the CSL, $d+id$ QSL and PQBT,  their Berry phase in the $\theta_1$ and $\theta_2$ plane exhibits different distribution. This difference is expected to be reflected as different temperature dependences of the thermal Hall effect, as it depends strongly on the momentum dependence of the Berry curvature\cite{PhysRevB.89.054420, PhysRevB.99.205157, Gao-thermal-hall-2020}. Though they have different distribtion of the Berry phase, we find that they have the same total Chern number 2 within our numerical errors, see Fig.~\ref{fig:topologe}.
To interpret the global topological structure, we calculate the GSDs of the two gapped chiral QSLs, namely the CSL and $d+id$ QSL. The detail calculations can been found in the Appendix A. We obtain $\mathrm{GSD} = 2$ for a U(1) CSL with $\psi=\frac{\pi}{2}$ and the $d + id$ QSL. So, both of two chiral states support semionic topological excitations\cite{zhang-prb-modular-2012}. Combining with the total Chern number, we infer that both of the two chiral spin liquids are Kalmeyer-Laughlin state with the same filling factor $\nu=\frac{1}{2}$\cite{Sheng-chern-num-2003, Zhang-tee-2011}. While, they are two different types of chiral spin liquids protected by different PSGs\cite{Bieri-psg-tri-2016}.

\section{\label{sec:conclusion}CONCLUSIONS}
In summary, we have investigated the interplay between the chiral interaction $J_\chi$ and four-site ring-exchange $J_4$ in the triangular $J_1$ Heisenberg model by use of the variational Monte Carlo  techniques. We map a detail  $J_\chi-J_4$ phase diagram,  in which the long-range magnetic $120^\circ$ order and five quantum disordered phases are identified, the latters include two chiral spin liquids with nontrivial Chern numbers and three achiral ones. The $J_4$ term alone induces QBT, nodal $d$-wave and U(1) SFS QSLs with its progressive increase. Among them, the nodal $d$-wave spin liquid is destabilized into the Z$_{2}$ $d + id$ spin liquid with any finite $J_\chi$  within our numerical calculation. And the U(1) spin liquid is more robust and turns to the chiral Z$_{2}$ $d + id$ spin liquid above critical $J_\chi$ values. In particular, we find a crossover region between the QBT spin liquid for $J_\chi=0$ and the Z$_{2}$ $d + id$ spin liquid once introducing the $J_\chi$ term. This proximate QBT state defined in this crossover region differs from the QBT spin liquid in that it acquires a nonzero hopping term and opens gaps at the quadratic touching point and two Dirac points. In this region, both the proximate QBT and $d + id$ spin liquid are stable solutions and compete with each other, while the former is energetically favored. With the further increase of $J_\chi$, the system will give up its preference of the proximate QBT state and enter into the Z$_{2}$ $d + id$ spin liquid with a more favorable energy. For the small $J_\chi$ and $J_4$, the $120^\circ$ magnetically ordered state is dominant. This phase is fully gapped but topological trivial (ground-state degeneracy GSD = 1). The $J_\chi$ term will eventually destabilize the $120^\circ$ order and prefers the U(1) CSL, and this U(1) CSL will also transit into the Z$_{2}$ $d + id$ spin liquid with the increase of the $J_4$ term. These results show that the Z$_{2}$ $d + id$ spin liquid occupies the largest region in the $J_\chi-J_4$ phase diagram due to the interplay between the  chiral interaction $J_\chi$ and four-spin term $J_4$.  Finally, we also show that U(1) CSL, proximate QBT state and the Z$_{2}$ $d + id$ QSL are topological nontrivial states with Chern number $C = 2$ and ground-state degeneracy GSD = 2.



\begin{acknowledgments}
We would like to thank Q.-H. Wang and Z.-X. Liu for many helpful and valuable discussions. This work was supported by National Key Projects for Research and Development of China (Grant No. 2021YFA1400400) and the National Natural Science Foundation of China (No. 92165205).
\end{acknowledgments}

\appendix

\section{Calculation of the ground-state degeneracy}


Firstly, we compact the triangular lattice on a torus, see Fig.~\ref{fig:torus}. In the thermodynamic limit, it does not cost any energy to insert a global $\pi$ flux into any of the two holes in the torus. In practice, this process is equivalent to changing periodic boundary condition of the mean-field Hamiltonian to anti-periodic one. From this, we can construct four mean-field ground states $|G_{\pm,\pm}^{\mathrm{mf}}\rangle$, where $\pm$ denotes the boundary conditions for the directions of $\vec{a}_{1,2}$, in detail, $+$ means periodic boundary condition and the $-$ means the anti one. Then, we apply a Gutzwiller projection to these mean-field ground states to obtain physical wave functions $|\Psi_{\pm,\pm}\rangle$ as the same as that in Sec.~\ref{sec2-model-method}. After these previous preparations, we can obtain the overlap (or density) matrix with the element given by
\begin{equation}
    \mathcal{O}_{ij} = \frac{\langle\Psi_i|\Psi_j\rangle}{\sqrt{\langle\Psi_i|\Psi_i\rangle\langle\Psi_j|\Psi_j\rangle}},
    \label{eq:overlap}
\end{equation}
where $i,j \in \{++, +-, -+, --\}$, and we emphasize that the normalization is necessary. Usually, $\mathcal{O}$ is a $4\times4$ matrix, but sometimes, we do not construct the whole four states under Gutzwiller projection, see the supplemental material in Ref. \cite{Wang-prl-2019}. In practice, the gap of a certain mean-field state is the bigger the better
for this progress to weaken the finite-size effect. After the construction,  we can diagonalize the overlap matrix to get four eigenvalues (or apply singular value decomposition), and the number of the significant eigenvalues is equal to the one of linearly independent states. It is just our final target, the ground-state degeneracy.



\nocite{*}

\bibliography{hlw-L12-v3}

\end{document}